\RequirePackage{ifpdf}
\ifpdf % We are running pdfTeX in pdf mode
\documentclass[pdftex]{sigma}
\else
\documentclass{sigma}
\fi

\newcommand{\nind}{{\mbox{\tiny $(N)$}}}
\newcommand{\dind}{{\mbox{\tiny $(2)$}}}
\newcommand{\dprime}{\prime}

\newcommand{\Ket}[1]{\left\vert#1\right\rangle}

\newcommand{\KetBra}[2]{\left\vert#1\right\rangle\left\langle#2\right\vert}

\newcommand{\hco}{{\hspace{0.3mm}\mathsf{h.c.}}}

\newcommand{\hbo}{H_{\mathrm{B}}}

\newcommand{\hcs}{H_{\mathrm{R}}}
\newcommand{\thcs}{\tilde{H}_{\mathrm{R}}}

\newcommand{\hlt}{H(\lambda; t)}
\newcommand{\ho}{H_0}
\newcommand{\hi}{H_{\diamond}}
\newcommand{\uo}{U_0}

\newcommand{\hilt}{\hi (\lambda ;t)}
\newcommand{\rotev}{T}

\newcommand{\zlt}{\hat{Z}(\lambda ;t)}

\newcommand{\ttt}{\mathfrak{t}}
\newcommand{\cltint}{\!\int_0^t\! C(\lambda ;\ttt)\, \de\ttt}
\newcommand{\rotham}{\tilde{H}}
\newcommand{\rothamlt}{\tilde{H}(\lambda ;t)}
\newcommand{\horder}[2]{\tilde{H}_{#1}(#2)}
\newcommand{\de}{\mathrm{d}}
\newcommand{\der}{\frac{\mathrm{d}\ }{\mathrm{d}t}}
\newcommand{\Ad}{\mathrm{Ad}}
\newcommand{\ad}{\mathrm{ad}}
\newcommand{\X}{\mathfrak{X}}

\newcommand{\ppar}{\lambda}

\newcommand{\rcpo}{\Omega}

\newcommand{\cpo}{\Omega}
\newcommand{\cpc}{g}

\begin{document}
\allowdisplaybreaks

\renewcommand{\PaperNumber}{050}

\FirstPageHeading

\ShortArticleName{Perturbative Treatment of the Evolution Operator
Associated with Raman Couplings}

\ArticleName{Perturbative Treatment of the Evolution Operator\\
Associated with Raman Couplings}

\Author{Benedetto MILITELLO~$^\dag$, Paolo ANIELLO~$^\ddag$ and
Antonino MESSINA~$^\dag$}

\AuthorNameForHeading{B. Militello, P. Aniello and A. Messina}

\Address{$^\dag\!$~INFM, MIUR and Dipartimento di Scienze Fisiche
ed
Astronomiche dell'Universit\`{a} di Palermo,\\
$\phantom{^\dag}\!$~Via Archiraf\/i, 36 - 90123 Palermo, Italy}
\EmailD{\href{mailto:bdmilite@fisica.unipa.it}{bdmilite@fisica.unipa.it},
\href{mailto:messina@fisica.unipa.it}{messina@fisica.unipa.it}}

\Address{$^\ddag\!$~Dipartimento di Scienze Fisiche
dell'Universit\`{a} di Napoli `Federico II' and INFN Sezione\\
$\phantom{^\ddag}\!$~di Napoli, Complesso Universitario di Monte
S.\ Angelo, Via Cintia - 80126 Napoli, Italy}
\EmailD{\href{mailto:paolo.aniello@na.infn.it}{paolo.aniello@na.infn.it}}

\ArticleDates{Received November 02, 2005, in f\/inal form April
13, 2006; Published online May 10, 2006}

\Abstract{A novel  perturbative treatment of the time evolution
operator of a quantum system is applied to the model describing a
Raman-driven trapped ion in order to obtain a~suitable
`ef\/fective model'. It is shown that the associated ef\/fective
Hamiltonian describes the system dynamics up to a certain
transformation which may be interpreted as a `dyna\-mi\-cal
dressing' of the ef\/fective model.}


\Classification{81Q15; 81V10; 81V45}

\section{Introduction}

Over the last few years more and more attention has been addressed
to the analysis of phy\-si\-cal nano-systems in order to realize
intriguing applications, for instance in the f\/ield of quantum
computation, and to investigate fundamental aspects of quantum
mechanics. One of the most {\it promising} physical contexts is
that of laser-driven trapped ions~(for a review see, for
instance~\mbox{\cite{Trap_Review,Leibfried}}).

An electromagnetic (e.m.)\ trap is a device which generates a
suitable e.m.\ f\/ield conf\/ining a~charged particle in a
f\/inite region of space. In particular, a Paul trap exploits an
inhomogeneous and time-dependent e.m.\ f\/ield which forces
a~charged particle to move approximately as a~harmonic oscillator.
The equilibrium point coincides with the center of the
trap~\cite{Ghosh}. Therefore, a trapped ion is describable as a
compound system made of a three-dimensional quantum harmonic
oscillator (representing the motion of the ion center of mass
inside the trap) and a~few-level system (associated with the
internal atomic state, i.e.\ with the `relevant electronic levels'
of the ion). Through the action of suitable driving laser
f\/ields, it is possible to coherently manipulate this system with
a high degree of accuracy. In particular, the possibility of
inducing couplings between the atomic degrees of freedom and the
ion center of mass motion has been experimentally demonstrated in
a very wide variety of
settings~\cite{Trap_Review,Leibfried,Ghosh}.

In many experimental situations, only two atomic states are {\it
effectively} involved in the dyna\-mics of a ion trap. In fact, such
two states are the only two {\it effectively} coupled by the laser
f\/ields driving the ion. Such ef\/fective couplings, for
technical reasons, are usually realized through a~third atomic
level. Precisely, the two {\it effective atomic levels} -- say
$\Ket{1}$ and $\Ket{2}$ -- are non-resonantly coupled to a third
level -- the {\it auxiliary level} $\Ket{3}$ -- and the respective
two `amounts of of\/f-resonance' (i.e.\ the two detunings, that is
the dif\/ferences between the atomic Bohr frequencies and the
corresponding laser frequencies) are chosen to be
equal~\cite{Trap_Raman}. The idea behind such a~coupling scheme --
the `Raman scheme' -- is that direct transitions from (and to)
level $\Ket{3}$ to (and from) the other two levels are forbidden
by the energy conservation, while two-photon processes bringing
from (and to) level $\Ket{1}$ to (and from) level $\Ket{2}$,
through the auxiliary level~$\Ket{3}$, are possible, and this
realizes an ef\/fective coupling $\Ket{1}\rightleftarrows\Ket{2}$.
We will show that such a reasoning is correct within a certain
approximation which will be clarif\/ied in the following (see
Sections~\ref{raman} and~\ref{conclusions}).\looseness=-1

In the present paper, we analyze the Raman coupling schemes
realized by laser-driven trapped ions deducing, by means of a {\it
rigorous} perturbative approach, the expressions of the relevant
ef\/fective couplings. To this aim, we exploit a recently
introduced perturbative method based on a suitable decomposition
of the time evolution operator associated with a quantum
Hamiltonian~\cite{Aniello_Method} (for the case of a
time-independent Hamiltonian, see also
\cite{Aniello_Method1,Aniello_Method2,Aniello_Method3}). In
particular, we investigate the problem of determining the {\it
effective Hamiltonian} -- i.e.\ the Hamiltonian describing the
ef\/fective couplings -- in the case where two or more Raman
coupling schemes are simultaneously active, so providing a
rigorous proof of the {\it additivity} of the ef\/fective
couplings.

We will show, moreover, that in the special case of a {\it single}
Raman coupling our result coincides with the result previously
obtained by means of a {\it time-independent} perturbative
app\-roach~\cite{Our_Paper}.

The paper is organized as follows. In Section~\ref{system}, we
describe the general form of the Hamiltonian associated with a
trapped ion Raman scheme. In Section~\ref{approach}, we introduce
the perturbative method which is the main tool of the paper. We
apply this method in Section~\ref{raman}, where we analyze the
dynamics of a double $\Lambda$ Raman scheme. Finally, in
Section~\ref{conclusions}, some conclusive remarks are drawn.

\section{The physical system}
\label{system}

The general form  of the quantum Hamiltonian of a trapped
three-level ion addressed by a set of laser beams coupling the
atomic level $\Ket{3}$ with the other two levels (with suitable
ion-laser detunings) is the following:
\begin{gather}
\label{DoubleRaman_Hamiltonian_Generic} H(t)=\ho+\hbo+\hcs(t),
\end{gather}
where
\begin{gather}
\ho=\sum_{l=1,2,3}\hbar\omega_l\hat{\sigma}_{ll} , \qquad
\hbo=\hbar\nu
\sum_{\alpha=x,y,z}\hat{a}_{\alpha}^{\dag}\hat{a}_{\alpha}^{\phantom{\dagger}},\nonumber
\\
\hcs(t)=\big(\hbar{\hat{\cpo}}_{13}(t)\,\hat{\sigma}_{13}+\hco\big)+
\big(\hbar{\hat{\cpo}}_{23}(t)
\hat{\sigma}_{23}+\hco\big),\label{HDefinitions}
\end{gather}
with $\hat{\sigma}_{lm}\equiv\KetBra{l}{m}$, $l,m=1,2,3$,
$\{\Ket{l}\}_{l=1}^3$ being the considered three atomic levels and
$\{\hbar\omega_l\}_{l=1}^3$ the corresponding energies
($\omega_l\neq\omega_r$, for $l\neq r$);
$\{\hat{a}_{\alpha}^{\phantom{\dagger}}:\,\alpha=x,y,z\}$ are the
annihilation operators associated with the center of mass harmonic
motion along the axes $x$, $y$, $z$, namely,
\begin{gather}\label{annihil_def}
\hat{a}_x^{\phantom{\dagger}}=\left(\frac{\mu\nu}{2\hbar}\right)^{1/2}\left(
\hat{x}+\frac{i}{\mu\nu}\,\hat{p}_x^{\phantom{\dagger}}\right),\quad\ldots,\quad
\hat{a}_z^{\phantom{\dagger}}=\left(\frac{\mu\nu}{2\hbar}\right)^{1/2}\left(
\hat{z}+\frac{i}{\mu\nu}\,\hat{p}_z^{\phantom{\dagger}}\right),
\end{gather}
which, of course, satisfy the well known bosonic commutation
relations
$\big[\hat{a}_\alpha^{\phantom{\dagger}},\hat{a}_\beta^{\phantom{\dagger}}\big]=
\big[\hat{a}_\alpha^{\dagger},\hat{a}_\beta^{\dagger}\big]=0$,
$\big[\hat{a}_\alpha^{\phantom{\dagger}},\hat{a}_\beta^{\dagger}\big]=\delta_{\alpha\beta}$,
for all $\alpha, \beta = x, y, z$. Without loss of generality, the
three harmonic oscillator frequencies have been taken to be equal:
$\nu_x=\nu_y=\nu_z\equiv\nu>0$; thus, we deal with a spherically
symmetric trap.

We have denoted by $t\mapsto\hat{\cpo}_{j3}(t)$, $j=1,2$, {\it
operator-valued} functions acting in the bosonic Hilbert space
(i.e.\ the Hilbert space associated with the vibrational degrees
of freedom); their specif\/ic structure is determined by the
specif\/ic laser conf\/iguration. For instance, for a single
`$\Lambda$~Raman coupling' -- involving two laser beams with
complex strengths (proportional to the laser amplitudes and to the
atomic dipole operators, and including the laser phases), wave
vectors and frequencies $g_{13}$, $\vec{k}_{13}$, $\omega_{13}$
and $g_{23}$, $\vec{k}_{23}$, $\omega_{23}$, respectively -- with
the lasers in a `travelling wave conf\/iguration', we have:
\begin{gather} \label{veryspecial}
\hat{\cpo}_{13}(t)=\hbar
g_{13}\,e^{-i\left(\vec{k}_{13}\cdot\,\vec{r}-\omega_{13}t\right)},\qquad
 \hat{\cpo}_{23}(t)=\hbar
g_{23}\,e^{-i\left(\vec{k}_{23}\cdot\,\vec{r}-\omega_{23}t\right)},
\end{gather}
where the laser frequencies are f\/ixed in such a way that the two
couplings share the same ion-laser detuning
\begin{gather}\nonumber
\Delta\equiv \omega_3 - \omega_1 -\omega_{13}=\omega_3- \omega_2
-\omega_{23}\neq 0,
\end{gather}
in order to allow the typical `two-photon processes'
$\Ket{1}\rightleftarrows\Ket{2}$ of the Raman scheme.
In~(\ref{veryspecial}) the vector operator $\vec{r} := (\hat{x},
\hat{y}, \hat{z})$ is the ion center of mass position operator and
its presence in the interaction Hamiltonian $\hcs(t)$ is
responsible for the interaction between atomic and vibrational
degrees of freedom. The link between this operator and the
annihilation and creation operators stems from
relations~(\ref{annihil_def}). With regard to the coef\/f\/icients
$g_{13}$ and $g_{23}$ appearing in the def\/inition of the
operators $\hat{\cpo}_{13}(t)$ and $\hat{\cpo}_{23}(t)$
respectively, we recall that they are given by
$g_{j3}:=-\hbar^{-1}\hspace{0.2mm}\vec{d}_{j3}\cdot\vec{E}_{j3}$,
where $\vec{E}_{j3}$ is the complex amplitude (i.e.\ including
information about the initial phase of the f\/ield) of the laser
f\/ield tuned near the $\Ket{j}\rightarrow\Ket{3}$ Bohr frequency,
while $\vec{d}_{j3}$ is the atomic dipole operator matrix element
involving the atomic states $\Ket{j}$ and $\Ket{3}$.

A relevant feature of the laser conf\/iguration specif\/ied by
relations~{(\ref{veryspecial})} is that, in this particular case,
the Hamiltonian $H(t)$ can be transformed, by passing to a
suitable rotating frame (i.e.\ by switching to an {\it ad hoc}
interaction picture), into a {\it time-independent} Hamiltonian,
which can be then treated by means of a {\it time-independent}
perturbative approach, see~\cite{Our_Paper}.

However, more complicated laser conf\/igurations are possible and
useful for various applications, and, in general, one cannot
f\/ind a simple rotating frame where the Hamiltonian of the system
becomes time-independent. Therefore, it will be convenient to
apply a {\it time-dependent} perturbative approach.

\section{The time-dependent perturbative approach}
\label{approach}

In order to study the class of dynamical problems associated with
a quantum Hamiltonian of the
form~{(\ref{DoubleRaman_Hamiltonian_Generic})}, one can fruitfully
exploit a time-dependent perturbative method based on a suitable
decomposition of the evolution operator~\cite{Aniello_Method}
which is a generalization of the classical Magnus
expansion~\cite{Magnus}.

Consider a quantum system whose Hamiltonian is made of two
components, the unperturbed energy operator $\ho$, and a
perturbation $\hilt$, in general time-dependent:
\begin{gather*} %\label{hamiltonian}
\hlt=\ho+\hilt .
\end{gather*}
We assume that $\lambda\mapsto\hilt$ is an analytic function of
the (real) {\it perturbative parameter} $\lambda$ ($\hi(0;t)=0$).

Introducing the evolution operator generated by the unperturbed
component $\ho$ -- namely, $\uo:=e^{-\frac{i}{\hbar}\ho t}$ -- and
the the evolution operator $\rotev(\lambda;t)$ associated with the
interaction picture Hamiltonian $\rotham(\lambda;t) :=
\uo(t)^\dagger \hilt \, \uo(t)$, it is possible to factorize the
total evolution opera\-tor $U(\lambda;t)$ of the system as
\begin{gather}\label{CompleteTimeEvol}
U(\lambda;t)= \uo(t)\, \rotev(\lambda;t).
\end{gather}

We can now consider the following exact decomposition of the
interaction picture evolution operator $\rotev(\lambda;t)$ as a
product of unitary operators:
\begin{gather} \label{gen1}
\rotev(\lambda;t) = \exp\left(-i\,Z(\lambda ;t)\right)\,
\exp\left(-i\int_{0}^t C(\lambda;\ttt)\,\de\ttt\right)
\exp\left(i\,Z(\lambda)\right),
\end{gather}
where $Z(\lambda ;t)$, $C(\lambda;t)$, $Z(\lambda)$ are
selfadjoint operators and
\begin{gather*}
Z(\lambda)\equiv Z(\lambda;0) ,\qquad Z(0;t)=C(0;t)=0,  \qquad
\forall\, t.
\end{gather*}
It is worth noting that no time-ordering operator is present in
(\ref{gen1}). Moreover, observe that in the special case where
$Z(\lambda ;t) = 0$, $\forall \, t$, we have the Magnus expansion,
provided that $C(\lambda;t)$ is regarded as the time derivative of
the Magnus unitary operator generator, see \cite{Magnus}; but, as
we will see below, decomposition~{(\ref{gen1})} is actually a
generalization of the Magnus expansion.

\subsection{Imposing a gauge condition}

It can be shown that the operators $C(\lambda ;t)$ and $Z(\lambda
;t)$ are not uniquely determined in decomposition~{(\ref{gen1})}:
in fact, there are inf\/inite possible solutions, namely,
solutions compatible with the general form of such decomposition.
A unique solution can be singled out by imposing an additional
`gauge condition'. A typical example is the case where the
interaction picture Hamiltonian $\rotham(\lambda;t)$ is a almost
periodic\footnote{A standard reference on almost periodic
functions is \cite{Amerio}.} operator-valued function of time, in
particular, an operator-valued {\it trigonometric polynomial} with
respect to the time variable. In this case, a remarkable gauge is
f\/ixed by the following tern of conditions
(see~\cite{Aniello_Method}):
\newcounter{xxx}
\begin{list}
{{\arabic{xxx})}}{\usecounter{xxx}}\itemsep=0pt \item $C(\lambda
;t)=C(\lambda;0)\equiv C(\lambda)$, $\forall \; t$;\label{uno}

\item the function $t\mapsto Z(\lambda ;t)$ satisf\/ies:
\label{due}
\begin{gather*}
\lim_{t\rightarrow\infty} t^{-1} Z(\lambda ;t)=0;
\end{gather*}

\item the {\it mean value} of the function $t\mapsto Z(\lambda
;t)$ is zero, namely: \label{tre}
\begin{gather*}
\lim_{\tau\rightarrow\infty}\frac{1}{\tau} \int_0^\tau Z(\lambda
;t)\, \de t =0 .
\end{gather*}
\end{list}

As it will be seen in Section~\ref{raman}, this is precisely the
case occurring in our applications, due to the fact that the
operators $\hat{\cpo}_{13}(t)$, $\hat{\cpo}_{23}(t)$ are indeed
trigonometric polynomials:
\[
\hat{\cpo}_{j3}(t)=\sum_{\kappa=1}^{\bar{\kappa}_j}
\hat{\cpo}_{j3}^{\kappa}\, e^{i\omega_j^{\kappa} t},\qquad
\bar{\kappa}_j\in\mathbb{N},\quad \big\{\omega_j^1,\ldots ,
\omega_j^{\bar{\kappa}_j}\big\}\subset\mathbb{R}, \quad j=1,2,
\]
where $\big\{\hat{\cpo}_j^1,\ldots ,
\hat{\cpo}_j^{\bar{\kappa}_j}\big\}_{j=1}^2$ are operators acting
on the bosonic degrees of freedom; see, for instance,
relations~{(\ref{veryspecial})}.

\subsection[Determination of $C(\lambda ;t)$ and $Z(\lambda ;t)$ up to a gauge condition]{Determination
of $\boldsymbol{C(\lambda ;t)}$ and $\boldsymbol{Z(\lambda ;t)}$
up to a gauge condition}

Using the formula -- reported, for instance, in~\cite{Wilcox} --
for the derivative of the exponential of an operator-valued
function $t\mapsto F(t)$, i.e.
\begin{gather*}
\der e^F = e^F \int_{0}^{1} \left(e^{-sF} \dot{F} e^{sF}\right)
\de s = \int_{0}^{1} \left(e^{sF} \dot{F} e^{-sF}\right)\de s \,
e^F ,
\end{gather*}
we can write the Schr\"odinger equation for the interaction
picture evolution operator as
\begin{gather}
   \rothamlt \rotev(\lambda ;t)    =    i\hbar\,\dot\rotev(\lambda ;t)
 =
\hbar e^{-iZ(\lambda ;t)}\int_{0}^{1}\left(e^{isZ(\lambda ;t)}
\dot{Z}(\lambda ;t) e^{-isZ(\lambda ;t)}\right)
\de s\, e^{-i\cltint} e^{iZ(\lambda)}  \nonumber \\
\phantom{\rothamlt \rotev(\lambda ;t)    =}{} + \hbar
e^{-iZ(\lambda ;t)}\!\!\int_{0}^{1}\!\!\! \left(e^{-is\cltint}
C(\lambda ;t) e^{is\cltint}\right)\! \de s\, e^{-i\cltint}
e^{iZ(\lambda)}.\!\! \label{preceding}
\end{gather}

Applying to each member of equation~{(\ref{preceding})} the
operator $e^{iZ(\lambda ;t)}$ on the left and the operator
$e^{-iZ(\lambda)}e^{i\cltint}$ on the right, we get the following
equation relating the operators $C(\lambda ;t)$ and $Z(\lambda
;t)$ with the interaction picture Hamiltonian:
\begin{gather} \label{prestart}
\Ad_{\exp(iZ(\lambda ;t))} \rothamlt = \hbar\int_0^1
\left(\Ad_{\exp(isZ(\lambda ;t))} \dot{Z}(\lambda ;t) +
\Ad_{\exp\left(-is\cltint\right)} C(\lambda ;t)\right)\de s.
\end{gather}
We recall that, given linear operators $\X$ (invertible) and $Y$,
$ \Ad_{\X} Y:=\X Y \X^{-1}$. If the opera\-tor~$\X$ is of the form
$\X=e^X$, we can use the well known relation
\begin{gather} \label{form}
\Ad_{\exp(X)} Y=\exp(\ad_X) Y =\sum_{m=0}^\infty
\frac{1}{m!}\ad_X^m Y,
\end{gather}
with $\ad_X^m$ denoting the $m$-th power
($\ad_X^0\equiv\mathrm{Id}$) of the adjoint super-operator $\ad_X$
def\/ined by $\ad_X\,Y:=[X,Y]$. Applying formula~{(\ref{form})} to
equation~{(\ref{prestart})}, and performing the integrals, we
obtain:
\begin{gather}
\sum_{m=0}^\infty \frac{i^m}{m!}\,\ad_{\zlt}^m\, \rothamlt    =
  \hbar
\sum_{m=0}^\infty \frac{i^m}{(m+1)!}\,\ad_{\zlt}^m\,
\dot{Z}(\lambda ;t)
\nonumber\\
\phantom{\sum_{m=0}^\infty \frac{i^m}{m!}\,\ad_{\zlt}^m\,
\rothamlt    = }{} +    \hbar \sum_{m=0}^\infty
\frac{(-i)^m}{(m+1)!}\,\ad_{\cltint}^m\,C(\lambda
;t).\label{eqcompl}
\end{gather}

Next, using the Taylor expansions of the operator-valued functions
$\lambda\mapsto\rothamlt$, $\lambda\mapsto Z(\lambda ;t)$ and
$\lambda\mapsto C(\lambda ;t)$ (recall that
$\rotham(0;t)=Z(0;t)=C(0;t)=0$), i.e.
\begin{gather}%\label{PParHamExpansion}
  \rothamlt=\sum_{n=1}^{\infty}\lambda^{n}\horder{n}{t},\qquad
  Z(\lambda ;t)=\sum_{n=1}^{\infty}\lambda^{n} Z_{n}(t) , \qquad
  C(\lambda ;t)=\sum_{n=1}^{\infty}\lambda^{n} C_{n}(t),\label{PParZCExpansion}
\end{gather}
from formula~(\ref{eqcompl}) we can determine (in a non-unique
way) -- order by order with respect to the perturbative parameter
$\lambda$ -- the operators $\{Z_n(t)\}_1^\infty$ and
$\{C_n(t)\}_1^\infty$.

\subsection[Solution corresponding to conditions~1)-3)]{Solution corresponding to conditions~\ref{uno})--\ref{tre})}

As anticipated, decomposition~{(\ref{gen1})} -- or, equivalently,
equation~{(\ref{eqcompl})} -- admits inf\/inite solutions.
Therefore, it is necessary to impose a gauge condition in order to
obtain a specif\/ic solution. For instance, as already mentioned,
the condition $Z(\lambda ;t) = 0$, $\forall\, t$, allows to obtain
a precise solution, that is the well known Magnus expansion.

In the following, we will use, instead, the gauge f\/ixed by the
tern of conditions~\ref{uno})--\ref{tre}). With these conditions,
one can single out a {\it unique} solution for the operators
$C(\lambda ;t)$ and $Z(\lambda ;t)$; precisely, we get an
inf\/inite set of equations that can be solved recursively, order
by order, for obtaining the operators
$\{C_n(t)=C_n(0)=:C_n\}_{n=1}^\infty$ and
$\{Z_n(t)\}_{n=1}^\infty$ that appear in the power
expansions~{(\ref{PParZCExpansion})}, up to an arbitrary order
$N$; specif\/ically, it turns out that

$1^{\mathrm{st}}\ \mbox{order:}$
\begin{gather}
C_1   = \lim_{\tau\rightarrow\infty}\frac{1}{\tau} \int_0^\tau
\hbar^{-1} \rotham_1(t)\, \de t  ,\nonumber
\\
Z_1(t)  =   \int_{0}^{t}{\hbar^{-1} \horder{1}{\ttt}\, \de \ttt}\
- C_1 t - \lim_{\tau\rightarrow\infty} \frac{1}{\tau}\int_0^\tau
\left( \int_0^t\left(\hbar^{-1}\rotham_1(\ttt)- C_1\right)
\de\ttt\right)\de t ; \label{forc1}
\end{gather}

$2^{\mathrm{nd}}\ \mbox{order:}$
\begin{gather}
 C_2   =
\lim_{\tau\rightarrow\infty}\frac{1}{\tau} \int_0^\tau
\left(\frac{i}{2}\;\ad_{Z_1(t)}\left(\hbar^{-1}\horder{1}{t}+
C_1\right) +\hbar^{-1}\horder{2}{t}\right)\de
t,\nonumber\\
Z_2(t)   =
\int_{0}^{t}\left(\frac{i}{2}\;\ad_{Z_1(\ttt)}\left(\hbar^{-1}\horder{1}{\ttt}+
C_1\right)
+\hbar^{-1}\horder{2}{\ttt}\right)\de \ttt - C_2 t \nonumber\\
\phantom{Z_2(t)   =}{} -
\lim_{\tau\rightarrow\infty}\frac{1}{\tau} \int_0^\tau \left(
\int_{0}^{t}\left(\frac{i}{2}\,\ad_{Z_1(\ttt)}\left(\hbar^{-1}\horder{1}{\ttt}+
C_1\right) +\hbar^{-1}\horder{2}{\ttt}-C_2\right)\de
\ttt\right)\de t ; \label{forc2}\\
 \cdots  \cdots  \cdots\cdots  \cdots
 \cdots  \cdots  \cdots\cdots  \cdots \cdots  \cdots  \cdots\cdots  \cdots
  \cdots  \cdots  \cdots\cdots  \cdots \cdots  \cdots  \cdots\cdots  \cdots  \cdots\nonumber
\end{gather}
Notice that the operators $C_1,Z_1(t),C_2,Z_2(t),\ldots$ are
indeed selfadjoint, coherently with our previous assumption.

As already mentioned, a typical case occurring in applications
(and, in particular, in the application considered in the present
paper) is the case where the coef\/f\/icients of the perturbative
expansion of the interaction picture Hamiltonian are
operator-valued trigonometric polynomials with respect to the time
variable (hence, almost periodic functions of time). In this case,
the gauge f\/ixed by conditions~\ref{uno})--\ref{tre}) is such
that the functions $\{t\mapsto Z_n(t)\}_{n=1}^\infty$ are
zero-mean-valued trigonometric polynomials; as a consequence, all
the `secular terms' are concentrated in the component of the
perturbative decomposition of the evolution operator which is
generated by the operators $\{C_n\}_{n=1}^\infty$, i.e.\ in the
one-parameter group of unitary operators $\{\exp(-i C(\lambda)
t)\}_{t\in\mathbb{R}}$.

\subsection[$N$-th order truncation of the perturbative
decomposition~(\ref{gen1})]{$\boldsymbol{N}$-th order truncation
of the perturbative decomposition~(\ref{gen1})} \label{truncation}

Once that the operators $C_1,Z_1(t),\ldots$ have been obtained
recursively up to a certain perturbative order $N\ge 1$, one can
write the following $N$-th order approximation of the interaction
picture evolution operator:
\begin{gather} \label{gen1_2}
\rotev(\ppar;t) \approx \exp\left(-iZ^\nind (\lambda ;t)\right)
\exp\left(-iC^\nind(\ppar) t\right)
\exp\left(iZ^\nind(\ppar)\right),
\end{gather}
with
\[
C^\nind(\ppar):=\sum_{n=1}^{N}\ppar^{n} C_{n}\qquad \mbox{and}
\qquad
 Z^\nind(\lambda ;t):=\sum_{n=1}^{N}\ppar^{n} Z_{n}(t).
\]
We stress that the $N$-th order truncation~{(\ref{gen1_2})}
preserves the fundamental unitary nature of the interaction
picture evolution operator $\rotev(\lambda ;t)$.

From formula~{(\ref{gen1_2})} we f\/ind that the overall evolution
operator of the system admits the following $N$-th order
approximation (recall relation~{(\ref{CompleteTimeEvol})}):
\begin{gather}
U(\lambda;t)    \approx    \uo(t) \exp\left(-iZ^\nind (\lambda
;t)\right) \exp\!\left(-iC^\nind(\ppar)\, t\right)
\exp\left(iZ^\nind(\ppar)\right) \nonumber\\
\phantom{U(\lambda;t)}{} =    \exp\big(-i\breve{Z}^\nind (\lambda
;t)\big) \uo(t) \exp\left(-iC^\nind(\ppar) t\right)\,
\exp\left(iZ^\nind(\ppar)\right),\label{exchange}
\end{gather}
where $\breve{Z}^\nind (\lambda ;t)= \uo(t) Z^\nind (\lambda ;t)
\uo(t)^\dagger$. From relation~{(\ref{exchange})} it follows that
\begin{gather*}
U(\lambda;t)\approx \exp\big(-i\breve{Z}^\nind (\lambda ;t)\big)
U_{\mathrm{eff}}^\nind(\lambda ;t)
\exp\left(iZ^\nind(\ppar)\right),
\end{gather*}
where $U_{\mathrm{eff}}^\nind(\lambda ;t)=\uo(t)
\exp\left(-iC^\nind(\ppar) t\right)$ can be regarded as the
evolution operator associated with the {\it effective Hamiltonian}
\begin{gather*}
H_{\mathrm{eff}}^\nind(\lambda ;t)= \ho + \hbar
\breve{C}^\nind(\lambda ;t)\qquad \mbox{with}\quad
\breve{C}^\nind(\lambda ;t):= e^{-\frac{i}{\hbar}\ho t}
C^\nind(\ppar)   e^{ \frac{i}{\hbar}\ho t} .
\end{gather*}

Thus, at the $N$-th perturbative order, $N\ge 1$, the total
evolution of the system can be decomposed into a `dynamical
dressing' -- i.e.\ the passage to a generalized interaction
picture generated by the time-dependent transformation
$\exp\big(-i\breve{Z}^\nind (\lambda ;t)\big)$, see
\cite{Aniello_Method} -- and the evolution generated by an
ef\/fective Hamiltonian $H_{\mathrm{eff}}^\nind(\lambda ;t)$
having the fundamental property that the corresponding interaction
picture Hamiltonian, with respect to the reference Hamiltonian
$\ho$ -- namely, $C^\nind(\ppar)$  -- is time-independent.

\section[Raman schemes: effective coupling]{Raman schemes: ef\/fective coupling}
\label{raman}

The perturbative approach described in the preceding section turns
out to be a powerful tool for studying Raman schemes and for
deducing the `ef\/fective couplings'
$\Ket{1}\rightleftarrows\Ket{2}$. For the sake of def\/initeness,
we consider the situation where two Raman setups are
simultaneously present. This case contains that of a single Raman
coupling as a special case\footnote{In Section~\ref{system}, we
have mentioned that the case of a~single Raman scheme can be
exceptionally treated by a~time-independent perturbative method.}.
Therefore, a direct comparison of the behaviors associated with
one or two Raman schemes can be given. It is worth noting that the
results obtained for two Raman schemes may be immediately
generalized to the case of several Raman couplings. Therefore, the
situation under scrutiny allows to illustrate all the relevant
conceptually remarkable aspects without introducing any cumbersome
notation.

The Schr\"odinger picture Hamiltonian describing two
simultaneously active Raman schemes is given by
equations~(\ref{DoubleRaman_Hamiltonian_Generic})
and~(\ref{HDefinitions}), now taking
\begin{gather*}
\hat{\rcpo}_{j3}(t) = \cpc_{j3} e^{-i\left(\vec{k}_{j3}\cdot
\vec{r}-\omega_{j3} t\right)}+ \cpc_{j3}^\dprime
e^{-i\left(\vec{k}_{j3}^\dprime\cdot\vec{r}-\omega_{j3}^\dprime
t\right)},\qquad j=1,2 ,
\end{gather*}
where $\cpc_{j3}$, $\omega_{j3}$, $\vec{k}_{j3}$, with $j=1,2$,
are the coupling constants, frequencies and wave vectors
associated with the f\/irst couple of Raman lasers, and
$\cpc_{j3}^\dprime$, $\omega_{j3}^\dprime$, $\vec{k}_{j3}^\dprime$
the analogous quantities for the second Raman scheme. We can
assume that $\cpc_{j3}\neq 0$ (and, of course, $\omega_{j3}\neq
0$, $\vec{k}_{j3}\neq 0$), $j=1,2$, so that the special case of a
{\it single} Raman coupling is recovered for $\cpc_{j3}^\dprime =
0$. As already mentioned, the operator $\vec{r}$ is the ion center
of mass position operator.

In order to generate `two-photon processes' involving levels
$\Ket{1}$ and $\Ket{2}$ (which is the main feature of the Raman
coupling), the ion-laser detunings are f\/ixed in such a way that
\begin{gather*}
(\omega_3-\omega_1)-\omega_{13}=
(\omega_3-\omega_2)-\omega_{23}\equiv    \Delta \neq 0,\\
(\omega_3-\omega_1)-\omega_{13}^\dprime=
(\omega_3-\omega_2)-\omega_{23}^\dprime   \equiv \Delta^\dprime
\neq 0 .
\end{gather*}
In the following, we will consider the regime where
$\Delta\neq\Delta^\dprime$; hence, the special case of a single
Raman coupling is recovered {\it only} for $\cpc_{j3}^\dprime =
0$, $j=1,2$. We can also assume, without loss of generality, that
$|\Delta|\ge |\Delta^\dprime|$, for $\cpc_{j3}^\dprime \neq 0$
(and, of course, $\omega_{j3}^\dprime\neq 0$,
$\vec{k}_{j3}^\dprime\neq 0$), $j=1,2$.

We will further suppose that the following {\it high detuning}
conditions are satisf\/ied:
\begin{gather} \label{hdetuning}
|\Delta| \gg |\cpc_{13}|, |\cpc_{23}|,\nu,\qquad (|\Delta|\ge)\
|\Delta^\dprime|\gg|\cpc_{13}^\dprime|,|\cpc_{23}^\dprime|.
\end{gather}

Passing to the interaction picture with respect to the reference
Hamiltonian $\ho$, the Schr\"o\-din\-ger picture Hamiltonian
$H(t)$ is transformed into the interaction picture Hamiltonian
\begin{gather*}
\rotham(t)=\hbo+\thcs(t),
\end{gather*}
where
\begin{gather*}
\thcs(t)=\big(\hbar\hat{\cpo}_{13}^{\widetilde{}}(t)\hat{\sigma}_{13}+\hco\big)+
\big(\hbar\hat{\cpo}_{23}^{\widetilde{}}(t)\hat{\sigma}_{23}+\hco\big),
\\
\hat{\cpo}_{j3}^{\widetilde{}}(t)= \cpc_{j3}
e^{-i\left(\vec{k}_{j3}\cdot \vec{r}+\Delta t\right)}+
\cpc_{j3}^\dprime
e^{-i\left(\vec{k}_{j3}^\dprime\cdot\vec{r}+\Delta^\dprime
t\right)},\qquad j=1,2 .
\end{gather*}
At this point, in order to put in evidence the `natural
perturbative parameter' of this model, it will be convenient to
introduce the {\it dimensionless} (interaction picture)
Hamiltonian $\mathfrak{H}(\lambda ;t)$ by setting:
\begin{gather*}
(\hbar\Delta)^{-1}\rotham(t)    =:    \mathfrak{H}(\lambda ;t) \\
\phantom{(\hbar\Delta)^{-1}\rotham(t)  }{ =    \lambda
\sum_{\alpha=x,y,z}\!\varkappa
\hat{a}_{\alpha}^{\dag}\hat{a}_{\alpha}^{\phantom{\dagger}} }
+\lambda \sum_{j=1,2}\! \left(\varkappa_{j3}
e^{-i\left(\vec{k}_{j3}\cdot\vec{r}+\Delta t\right)}
\hat{\sigma}_{j3} + \varkappa_{j3}^\dprime
e^{-i\left(\vec{k}_{j3}^\dprime\cdot\vec{r}+\Delta^{\dprime}t\right)}
\hat{\sigma}_{j3} +\hco\right),
\end{gather*}
where $\lambda$ is the dimensionless (real) perturbative parameter
def\/ined by
\begin{gather*}
\lambda := \frac{g}{\Delta} ,\qquad
g\equiv\mathrm{max}\{\nu,|\cpc_{13}|, |\cpc_{23}|,
|\cpc_{13}^\dprime|,|\cpc_{23}^\dprime|\},
\\
\mbox{and} \qquad \varkappa\equiv\nu/g,\qquad \varkappa_{j3}\equiv
\cpc_{j3}/g,\qquad \varkappa_{j3}^\dprime\equiv
\cpc_{j3}^\dprime/g,\qquad j=1,2.
\end{gather*}
We notice explicitly that $0<\varkappa\le1$, $|\varkappa_{j3}|\le
1$, $|\varkappa_{j3}^\dprime|\le 1$, $j=1,2$, and, due to
conditions~{(\ref{hdetuning})}, we have that $|\lambda|\ll 1$;
hence, $\lambda$ is indeed a `good perturbative parameter'
(compare with the time-independent perturbative approach used
in~\cite{Our_Paper} for the single Raman coupling).

Next, applying formulae (\ref{forc1}) and (\ref{forc2}) to the
interaction picture Hamiltonian $\hbar\Delta\,
\mathfrak{H}(\lambda ;t)$, we easily f\/ind the following
expressions for the operators $C_1$, $C_2$ (which, as we have
seen, determine the second order ef\/fective Hamiltonian):
\begin{gather} \label{C1Expression}
\ppar\, C_1    =    \hbar^{-1} \hbo =\nu\sum_{\alpha=x,y,z}
\hat{a}_{\alpha}^{\dag}\hat{a}_{\alpha}^{\phantom{\dagger}},
\\
\ppar^2 C_2 =   \breve{\omega}_1 \hat{\sigma}_{11} +
\breve{\omega}_2 \hat{\sigma}_{22} + \breve{\omega}_3
\hat{\sigma}_{33} +   \left(
    \left(\cpc_{12}
    e^{-i\vec{k}_{12}\cdot\vec{r}}+
 \cpc_{12}^\dprime
    e^{-i\vec{k}_{12}^\dprime\cdot \vec{r}}
    \right)\hat{\sigma}_{12}
    + \hco
  \right),\label{C2Expression}
\end{gather}
where we have set
\begin{gather*}
\breve{\omega}_j    :=    - \frac{|\cpc_{j3}|^2}{\Delta} -
\frac{|\cpc_{j3}^\dprime|^2}{\Delta^\dprime} ,\qquad j=1,2, \qquad
\breve{\omega}_3    :=
\frac{|\cpc_{13}|^2+|\cpc_{23}|^2}{\Delta}+
    \frac{|\cpc_{13}^\dprime|^2+|\cpc_{23}^\dprime|^2}{\Delta^\dprime},
\end{gather*}
and
\begin{gather} \label{effconst}
\cpc_{12}:=\frac{\cpc_{13}\cpc_{32}}{\Delta} ,   \qquad
\cpc_{12}^\dprime :=
\frac{\cpc_{13}^\dprime\cpc_{32}^\dprime}{\Delta^\dprime} ,\qquad
\vec{k}_{12}:=\vec{k}_{13} -\vec{k}_{23} , \qquad
\vec{k}_{12}^\dprime :=
\vec{k}_{13}^\dprime-\,\vec{k}_{23}^\dprime ,
\end{gather}
with $\cpc_{3j}^{\phantom{\ast}}\equiv\cpc_{j3}^{\ast}$,
$\cpc_{3j}^\dprime\equiv\cpc_{j3}^{\dprime \ast}$, $j=1,2$.

Now, according to what we have shown in
Subsection~\ref{truncation}, the {\it complete} temporal evolution
of the Raman-driven trapped ion, evaluated within the second order
in the perturbative parameter, can be written as
\begin{gather}
U(\ppar, t)    \approx e^{-\frac{i}{\hbar}\ho t} \exp(-i
Z^\dind(\ppar; t)) e^{-i(\ppar C_1 + \ppar^2
  C_2)t}
\exp(i Z^\dind(\ppar))\nonumber\\
\phantom{U(\ppar, t)}{} = \exp(-i \breve{Z}^\dind(\ppar; t))
e^{-\frac{i}{\hbar}\ho t} e^{-i(\ppar C_1 + \ppar^2 C_2)t} \exp(i
Z^\dind(\ppar)), \label{CompleteTwoRamanTimeEvol}
\end{gather}
where
%\begin{gather*}%\label{DefZetaTilde}
$\breve{Z}^\dind(\ppar; t):=e^{-\frac{i}{\hbar}\ho t}
Z^\dind(\ppar; t) e^{\frac{i}{\hbar}\ho t}
$
%\end{gather*}
is an operator-valued function of time, which, as a consequence of
of conditions~\ref{uno})--\ref{tre}), is a~zero-mean-valued
trigonometric polynomial, hence contains only `oscillatory terms';
its explicit form, however, is not relevant for the purposes of
the present paper and will be omitted.

Thus, as relation~{(\ref{CompleteTwoRamanTimeEvol})} shows, up to
the `dynamical dressing' generated by the time-dependent unitary
transformation $\exp(-i \breve{Z}(\ppar; t))\approx \exp(-i
\breve{Z}^\dind(\ppar;t))$, the temporal evolution of the system
is described, at the second perturbative order, by the evolution
operator $e^{-\frac{i}{\hbar}\ho t} e^{-i(\ppar C_1 + \ppar^2
C_2)t}$. Such an evolution operator may be regarded as a {\it
Schr\"odinger picture evolutor} factorized into the the
unperturbed evolution $e^{-\frac{i}{\hbar}\ho t}$ and the
interaction picture evolution $e^{-i(\ppar\, C_1 + \ppar^2
C_2)t}$, which is the one-parameter group of unitary operators
generated by the time-independent `interaction picture
Hamiltonian' $\ppar\, C_1 + \ppar^2 C_2$.

The corresponding Schr\"odinger picture Hamiltonian is then given
by
\begin{gather*}
H_{\mathrm{eff}}^\dind(t):= \ho + e^{-\frac{i}{\hbar}\ho
t}\left(\ppar\, C_1 + \ppar^2 C_2\right) e^{\frac{i}{\hbar}\ho t},
\end{gather*}
and -- as it is easily checked using
formulae~{(\ref{C1Expression})} and~{(\ref{C2Expression})} -- it
can be decomposed by means of the complementary orthogonal
projectors $\hat{P}_{12}\equiv\hat{\sigma}_{11}+\hat{\sigma}_{22}$
and $\hat{\sigma}_{33}$
($\hat{P}_{12}+\hat{\sigma}_{33}=\mathrm{Id}$,
$[\hat{P}_{12},\hat{\sigma}_{33}]=0$):
\[
H_{\mathrm{eff}}^\dind(t) = H_{\mathrm{eff}}^\dind(t)\hat{P}_{12}
+ H_{\mathrm{eff}}^\dind(t)\hat{\sigma}_{33} ,\qquad
[H_{\mathrm{eff}}^\dind(t),\hat{P}_{12}]=[H_{\mathrm{eff}}^\dind(t),\hat{\sigma}_{33}]=0,
\]
where
\begin{gather}
H_{\mathrm{eff}}^\dind(t)\hat{P}_{12}    = \hbar\nu
\sum_{\alpha=x,y,z}\hat{a}_\alpha^\dagger
\hat{a}_\alpha^{\phantom{\dagger}} \hat{P}_{12} +
\hbar(\omega_1+\breve{\omega}_1)\hat{\sigma}_{11}
+\hbar(\omega_2+\breve{\omega}_2)\hat{\sigma}_{22}
\nonumber \\
\phantom{H_{\mathrm{eff}}^\dind(t)\hat{P}_{12}    =}{} +
\left(\left(\hbar
g_{12}e^{-i\left(\vec{k}_{12}\cdot\,\vec{r}-\omega_{12}t\right)}
+\hbar g_{12}^\dprime e^{-i\left(\vec{k}_{12}^\dprime\cdot
\vec{r}-\omega_{12}^\dprime t\right)}
\right)\hat{\sigma}_{12}+ \hco \right), \label{decoupled1}\\
H_{\mathrm{eff}}^\dind(t) \hat{\sigma}_{33}    = \hbar\nu \!\!
\sum_{\alpha=x,y,z}  \hat{a}_\alpha^\dagger
\hat{a}_\alpha^{\phantom{\dagger}}\hat{\sigma}_{33}+
\hbar(\omega_3+\breve{\omega}_3)\hat{\sigma}_{33} ,\nonumber
%\label{decoupled2}
\end{gather}
with the ef\/fective frequencies $\omega_{12}$,
$\omega_{12}^\dprime$ def\/ined by
\begin{gather} \label{efffreq}
\omega_{12}:=\omega_2-\omega_1=\omega_{13}-\omega_{23} ,\qquad
\omega_{12}^\dprime:=\omega_2^\dprime-\omega_1^\dprime=\omega_{13}^\dprime-\omega_{23}^\dprime.
\end{gather}

The result found has a simple and transparent interpretation. In
fact, the ef\/fective Hamiltonian $H_{\mathrm{eff}}^\dind(t)$ is
the sum of two completely {\it decoupled} Hamiltonians, `living'
respectively in the ranges of the orthogonal projectors
$\hat{P}_{12}$ and $\hat{\sigma}_{33}$. It is worth noting the
remarkable fact that {\it the Hamiltonian
$H_{\mathrm{eff}}^\dind(t)\,\hat{P}_{12}$ can be regarded as the
standard Hamiltonian of a trapped two-level ion in interaction
with a couple of laser fields characterized, respectively, by the
following effective parameters}: $g_{12}$, $\vec{k}_{12}$,
$\omega_{12}$ and $g_{12}^\dprime$, $\vec{k}_{12}^\dprime$,
$\omega_{12}^\dprime$ (see formulae~{(\ref{effconst})}
and~{(\ref{efffreq})}).

\section{Conclusions}
\label{conclusions}

We can eventually draw the following conclusions:
\begin{itemize}\vspace{-2mm}\itemsep=0pt

\item the quantum Hamiltonian describing the physical system of a
trapped three-level ion in interaction with a set of laser beams
generating Raman couplings can be successfully treated by means of
a suitable time-dependent perturbative approach;

\item the result of this treatment, in the special case of a
single Raman coupling, coincides with the result obtainable  by
means of a time-independent perturbative approach,
see~\cite{Our_Paper};

\item at the second perturbative order (hence, with a high degree
of accuracy), the dynamics of the system is given, up to a
`dynamical dressing', by the dynamics associated with an
ef\/fective Hamiltonian describing two decoupled subsystems: an
ef\/fective laser-driven trapped two-level ion and a simple
harmonic oscillator; this result shows also the additivity of the
ef\/fective couplings: indeed, the ef\/fective Hamiltonian
associated with a certain set of Raman couplings is characterized
by a coupling term which is the {\it sum} of the single
ef\/fective couplings (see formula~{(\ref{decoupled1})});

\item although a detailed analysis of this point is beyond the aim
of the present paper, it is worth mentioning the fact that the
behavior of {\it certain} experimentally observable quantities, in
correspondence to {\it certain} initial conditions of the system,
is rather well reproduced by the ef\/fective dynamics only --
i.e.\ neglecting the ef\/fect of the dynamical dressing -- as
a~consequence of the temporal coarse-graining introduced by the
experimental apparatus (which has, unavoidably, a f\/ixed temporal
resolution); nevertheless, due to the presence of the dynamical
dressing, in addition to the standard Rabi oscillations between
the ef\/fective le\-vels~$\Ket{1}$ and $\Ket{2}$, also fast
transitions coupling these two levels with the auxiliary
level~$\Ket{3}$ take place; then, if the auxiliary level is an
excited level with non-negligible decay rate towards (at least one
of the) levels~$\Ket{1}$ and $\Ket{2}$, the fast transitions to
the auxiliary level, composed with decays, gradually injects
decoherence into the ef\/fective coherent cycle involving
 levels~$\Ket{1}$ and $\Ket{2}$ (see~\cite{Our_Paper}), result which
is actually in agreement both with experimental
observations~\cite{Nist_PRL,Leibfried1997} and with numerical
simulations on the basis of a phenomenological master
equation~\cite{VogelRaman}.\vspace{-2mm}

\end{itemize}

A detailed study of the experimentally observable implications of
the dynamical dressing, especially with regard to the decoherence
ef\/fects, is still work in progress~\cite{progress}, and it seems
to raise very intriguing issues both on the theoretical and on the
experimental side.

\subsection*{Acknowledgements}

The main results of the paper where presented by one of us (B.M.)
during the international conference {\it Symmetry in Nonlinear
Mathematical Physics} (June 20--26, 2005, Kyiv). The authors wish
to thank the kind organizers.

\LastPageEnding

\end{document}